\documentstyle[12pt,epsf,epsfig]{article}
\newcommand {\bi} {\bibitem}
\newcommand {\be} {\begin{equation}}
\newcommand {\bea} {\begin{eqnarray} \nonumber }
\newcommand {\ee} {\end{equation}}
\newcommand {\eea} {\end{eqnarray}}

\newcommand {\si} {\sigma}

\newcommand {\ga} {\gamma}

\newcommand {\al} {\alpha}

\newcommand {\lan} {\langle}
\newcommand {\ran} {\rangle}

\newcommand {\cS}  {{\cal S}}

\newcommand {\cF} {{\cal F}}

\def \form#1 {eq. (\ref{#1}) }
\def \parziale#1#2  {{\partial {#1} \over \partial {#2}}}

\begin{document}

\title{On the Approach to the Equilibrium and the 
Equilibrium Properties of a Glass-Forming Model}
\author{Barbara Coluzzi and Giorgio Parisi\\[0.5em]
{\small Dipartimento di Fisica, Universit\`a di Roma} 
{\small {\em La Sapienza} }\\
{\small INFN, sezione  di Roma} {\small {\em La Sapienza} }\\
{\small   \ \  Piazzale A. Moro 5, 00185 Roma (Italy)}\\[0.3em]
{\small   \tt coluzzi@chimera.roma1.infn.it} \\
{\small \tt giorgio.parisi@roma1.infn.it}\\[0.5em]
}
\maketitle

\begin{abstract}
In this note we apply some theoretical predictions that arise in the mean field
framework for a large class of infinite range models to  
structural glasses and we present a first comparison of these predictions 
with numerical results. 
\end{abstract}

\newpage

\section{Introduction}

\noindent
In recent years some theoretical progresses in our understanding of glasses 
have been achieved by 
comparing the results obtained for soluble models of generalized spin glasses 
\cite{EA,mpv}
with structural 
glasses properties under the assumption that the phase space of 
the two systems is similar.  These progresses are also due 
to the use of the replica method and of related concepts as replica symmetry 
breaking, coupled replicas, dynamic transitions \ldots

In the nutshell many of the ideas of this replica approach are already present 
in the original 
papers of Gibbs and Di Marzio \cite{gdi}. However, the use of the whole 
panoply of tools developed 
in the study of spin glass models allow us to put these ideas in a much 
sharper form and to test 
them in numerical (and possibly real) experiments.  Moreover the comparison 
of structural glasses 
and generalized spin glasses, introduced in ref.\cite{kirtir}, has proven 
extremely useful and the 
conjecture that the two models have similar energy landscapes has been a 
fruitful starting point.

In this note we will not discuss the theoretical basis under which this 
scenario has been derived but 
we will concentrate the attention on the physical picture and 
on the consequences 
of these predictions 
on the statistical properties of relative small samples.

The basic assumption is that, at sufficiently low temperatures but 
still in the liquid 
phase, the system is almost always trapped for a long time in one of the 
exponentially large number 
of local minima of the free energy.  The number ($\cal N$) of these local 
minima is related to the 
configurational entropy or complexity $\Sigma$ by 
\be
{\cal N}\approx\exp(N\Sigma(T)),
\ee
$N$ being the
number of particles and the previous formula being asymptotic for large $N$.  
The total entropy 
density $S$ is the sum of two contributions: the entropy density of each 
minimum and the complexity.
This description is valid for $T_{K}<T<T_{D}$.  The complexity 
\cite{mon} (which starts 
from a non zero value at 
$T_{D}$) is supposed to vanish linearly at a lower temperature, (i.e.  at the 
temperature $T_K$) 
where the height of the typical barriers becomes infinite.  The correlation 
time diverges at $T_K$, 
and one can argue in favour of a Vogel-Fulcher law (e.g.  
$\tau \propto \exp(-A (T-T_{K})^{-\nu})$, where $\nu \simeq 1$).

This scenario is implemented in a large class of infinite range models 
\cite{kirtir}, where a more detailed picture of the phase space of the system 
is obtained.  
Indeed these infinite range models are soluble because the 
appropriate mean field 
theory is exact.  The study of the mean field theory for these models is not 
a trivial task: the 
phase space structure of the configurations with low energy is quite complex 
and this is shown up 
in a rather interesting behaviour of the system both at equilibrium and in 
its approach to 
equilibrium.  The existence of these complex structures implies the need of 
using modern tools (like 
the replica formalism) to study the properties of these systems.

The aim of this note is to spell out some of the theoretical predictions that 
are obtained in the 
mean field framework for these infinite range models, to apply them (with the 
appropriate 
modifications) to the case of structural glasses and to present a first 
comparison of these 
predictions with numerical simulations.

\section{A mini theoretical review}

\noindent
The main hypothesis of the Gibbs-Di Marzio's approach is that at 
low temperatures 
the system quite always 
stays close to a minimum of the free energy and that the long time dynamics 
is dominated by the time 
needed to escape from one valley to another.

As noticed in ref. \cite{kirtir} this scenario is implemented in 
long range models in which detailed computations can be done and a more 
precise picture can be obtained \cite{pa1,pa2}.  

Let us consider for definiteness a spin model where the local variables are 
spins ($\si(i)$, $i=1 
\dots N$).  They could be either Ising spins $\si(i)=\pm 1$ or spherical 
spins, i.e.  real 
variables which satisfy the constraint
\be
\frac{\sum_{i=1}^{N}\si(i)^{2}}{N}=1.
\ee
The Hamiltonian $H(\si)$ has a form that we do not need to specify here.  
There are many different models with quite different Hamiltonians which share 
the properties that we are going to describe. In the following we will 
consider  a {\sl finite} large 
system and only at the end we will send the value of $N$ to $\infty$.

We suppose that the phase space can be broken into many valleys separated by 
high mountains. In 
other words, the free energy (or the energy at low temperature) has 
many minima 
and the free energy 
barrier which we have to cross for going from one state to another 
is 
quite large. In the 
infinite range approximation it will be proportional to $N$.  We can consider 
all the local minima 
of the energy as functions of the configuration and we can associate a valley 
to each of them at zero 
temperature.  At higher temperature we have to consider the minima of the 
free energy and when increasing 
the temperature the valleys will disappear. 

We could also define the valleys in a dynamical way as regions of 
configuration space in which the system remains trapped for a large time.  
This definition is similar to the previous one if we assume that the large 
time behaviour of the dynamics at low temperature is dominated by activated 
processes which correspond to the crossing of high free energy barriers.

Each valley (which we label by $\al$) may be characterized by a magnetization
\be
m_{\al}(i)=\lan\si(i)\ran_{\al},
\ee
where $\lan \cdot \ran_{\al}$ is the statistical expectation value restricted 
to the $\al$-valley. One can define a free energy $F[m]$, which is a function 
of all the local magnetizations, whose functional form depends on the system. 
The valleys are local minima of this free energy.

\subsection{Equilibrium properties}
\noindent
One finds in a very large class of models the following behaviour for the 
equilibrium properties depending on the temperature \cite{mon}-
\cite{cagiapa}:
\begin{itemize}
\item For temperatures higher than $T_{m}$ there is only one minimum of the
free energy $F[m]$ at 
$m(i)=0 \: \forall i$.  
In this case the total free energy of the system $\cF$ is given by 
$\cF=F[0]$.
\item At temperatures lower than $T_{m}$ there is an exponentially large 
number of minima. The contribution of these minima to the partition function 
can be estimated by
\be
Z_{m}=\int df \exp(-N\beta f +N\Sigma(f,\beta))\approx \exp(-N\beta f^{*} 
+N\Sigma(f^{*},\beta)),
\ee
where $f$ is the free energy density ($Nf=\cF$), $\exp (N\Sigma(f,\beta))$ is 
the number of minima of the free energy density $f$ and $f^{*}$ is the value 
of 
$f$ which maximizes the exponent, that is a function of $\beta$.
\end{itemize}

Below $T_{m}$ we can distinguish three regions:
\begin{itemize}
\item
For $T>T_{D}$ the contribution to the partition function of the non trivial 
minima can be neglected and the free energy is still given by $F[0]$.
\item For $T_{D}>T>T_{K}$ the contribution to the partition function of the 
non trivial minima is dominant.  The number of minima which dominate the 
partition function is exponentially large and the total entropy of the system 
is given by
\be
\cS= S_{m}+N\Sigma(f^{*},\beta),
\ee
where $S_{m}$ is the contribution to the total entropy of one minimum and 
$\Sigma(f^{*},\beta)>0$.

The magnetization averaged over all the minima is given by 
$m(i)=\lan\si(i)\ran=\sum_{\al}w_{\al}\lan\si(i)\ran_{\al}$, where 
$w_{\al}\propto \exp(-\beta F_{\al})$ is proportional to the contribution of 
the $\al$-minimum to the partition function ($\sum_{\al}w_{\al}=1$).

In this region the magnetization averaged over all the minima is zero and 
the total free energy which is dominated by the contribution of all the non 
trivial minima is still given by $F[0]$.  The free energy and all the other 
static equilibrium properties (quantities?) are fully regular at $T_{D}$.

\item
For $T<T_{K}$, $\Sigma(f^{*},\beta)=0$.  For $T$ slightly 
larger than $T_{K}$ one finds $\Sigma(f^{*},\beta) \propto (T-T_{K})$, 
so that the complexity (and consequently the entropy) has a discontinuity in 
the 
temperature derivative at $T=T_{K}$, which is the transition point from a 
thermodynamic point of view.  Here the entropy becomes equal to the 
contribution of a single minimum and this temperature may be identified with 
the Kauzmann temperature.
\end{itemize}

The region $T<T_{K}$ can be characterized by a peculiar behaviour.  The 
partition function is dominated by those states which have the minimum free 
energy and there are some minima for which quantities $w_{\al}$ remain of 
order 1.

In the previous description we have neglected the possibility of 
crystallization, i.e.  the formation of a highly ordered state which leads to 
a first order transition. If we consider this possibility we have to 
distinguish two cases, systems with quenched disorder in the Hamiltonian and 
systems without disorder (often with a translation invariant Hamiltonian).
\begin{itemize}
\item There are systems whose Hamiltonian contains quenched disorder.  
A typical example would be:
\be
H=\sum_{i,k,l}J(i,k,l)\si(i) \si(k)\si(l), \ee
where the variables $J$ are random (e.g. Gaussian).

Another example (in which the mean field approximation is non exact) would 
be a system in which the 
particles interact not only with themselves but also with an external, fixed, 
random potential.  It 
is clear that if the external potential is strong enough the free energy of 
the crystal phase may 
become quite large, while that of the glassy phase may be much less 
affected.
\item 
Other systems may not contain quenched variables in the Hamiltonian.  A typical
example \cite{mapari} is:
\be
H=\frac{1}{\sqrt N} \sum_{i,k=1}^{N}
\sin \left ( \frac{2 \pi i k}{N} \right ) (\si(i)-\si(k))^{2}
\hspace{.3cm} \mbox{with} \hspace{.3cm} \si(i)=\pm1. 
\ee
Another example  would be a system in which the 
particles interact only with themselves with a given potential.
\end{itemize}
The two categories of models seem rather different one from another; however 
it has been noticed 
that in the mean field approximation systems belonging to the two categories 
behave in a quite 
similar way, with the only difference that systems without quenched disorder 
may crystallize.

While this kind of results can be proved in the long range models, their 
validity for short range 
models (like structural glasses) remains an open question.  In this note we 
would like to present 
numerical evidence for the correctness of these ideas also for 
structural glasses.  Before doing so we 
will examine in detail the predictions of the mean field approach.

\subsection{Equilibrium properties at low temperature}
For simplicity let us consider the predictions that would follow from the 
application of the 
replica approach to a system of $N$ particles interacting with a given 
Hamiltonian in a box.

We will consider for simplicity a system without any type of symmetry (the box 
is not symmetric and 
the particles interact with the walls, periodic boundary conditions are not 
used so that there is no
translational invariance).  We suppose that for a given value of $N$ the 
Hamiltonian has many minima which we label by $\al$.

The partition function can thus written as
\bea
Z(\beta,N) \simeq \sum_{\al}\exp (-\beta F_{\al}(\beta,N))\\
F_{\al}(\beta,N) =E_{\al,N} -T S(\al,N),
\eea
where for simplicity we label the configurations in increasing free 
energy order. For small 
temperatures the entropic contribution to the free energy will be as usual 
negligible.  It is quite 
evident that for a finite system the partition function is dominated by the 
lowest energy 
configurations in the limit of zero temperature, and we will suppose that this 
property persists in 
the infinite volume limit.

For different values of $N$ we will have quite different values of the free 
energies.  The lowest 
free energy states will have free energy differences of order 1, so that by 
adding a single particle 
(i.e.  going from $N$ to $N+1$) we can strongly change the values of the 
$F_{\al}(\beta,N)$.  If we 
label by $F_{0}$ the lowest free energy we expect that the quantities 
$F_{\al}(\beta,N)- 
F_{0}(\beta,N)$ remain of order 1 when $N$ goes to infinity, but they do not 
go to any limit (they have no $N \rightarrow \infty$ limit?) because 
they change with $N$.  In this case it is natural to introduce a probability 
distribution for the 
differences in free energies, which tell us the probability of finding 
a given value of 
$F_{\al}(\beta,N)- F_{0}(\beta,N)$ if we chose randomly (albeit large) the 
value of $N$.  In other 
words if the value of the free energy changes strongly with $N$ it is 
convenient to introduce the probability distribution of their values.

The precise construction is the following: for any large value of $N$ we 
introduce a reference 
free energy $F_{R}(\beta,N)$ such as that, in the whole region 
$F-F_{R}(\beta,N)<<N$, the probability of 
finding a minimum of free energy $F$ is given by
\be
P(F) =\exp (\beta (x(\beta) (F-F_{R}(\beta,N))),
\ee
where the quantity $x(\beta)$ parameterizes the probability distribution.

The condition that the free energy is approximately given by $F_{R}$ implies 
that the integral
\be
\int_{F_{R}(\beta,N)}^{\infty}P(F) \exp(-\beta (F-F_{R}(\beta,N))
\ee
is convergent and therefore $x(\beta)\le 1$.  

The probability of finding a configuration in the minimum labeled by $\al$ is
\be
w_{\al}\propto\exp(-\beta F_{\al}),
\ee
where the normalization condition  
\be
\sum_{\al}w_{\al}=1
\ee
is satisfied.

In many models $x$ becomes equal to 
1 at the temperature $T_{K}$. At low temperatures
$x$ is proportional to the temperature (it would be exactly proportional to 
the temperature if we 
neglect the entropic contribution).

An important property of the model is how different the various minima are.  
The simplest hypothesis 
is that they are as different as possible, i.e.  the correlations among the 
particles in one minimum 
and in another minimum (among those of lowest energy) are zero.  For example 
the probability of 
finding two particles in two different minima at a given distance $r$ does 
not depend on the distance:
\be
P_{\al,\gamma}(r)=\sum_{i,k}\delta \left( x_{\al}(i)-x_{\ga}(k) 
-r \right ) =\rho^{2}
\ee
This hypothesis is usually called one step replica symmetry breaking.  More 
complicated distributions 
of the distances are discussed in the literature.

If there are symmetries the situation becomes slightly more complex. 
For example 
in a translational 
invariant system if $x(i)$ are the coordinates of a minimum, 
$x(i)+\delta$ are the 
coordinates of another minimum.  It is therefore useful to consider all the 
minima which are 
related by a symmetry transformation as a single minimum.

\subsection{The probability distribution of the distance}
\noindent
The properties of the system may be sharpened by introducing an appropriately
defined distance $d$ between configurations of particles and looking at  
the corresponding equilibrium probability distribution:
\be
P_{N}(d)=\sum_{\al,\ga}w_{\al}w_{\ga} \delta(d-d_{\al \ga}).
\ee

In the limit of large $N$ we have that
\be
P_{N}(d)= a_{N}\delta(d_{0}-d)+b_{N}\delta(d_{1}-d),
\ee
where $a_{N}+b_{N}=1$.  Of course for a finite system the delta function will 
be smoothed.

The function $a_{N}$ represents the probability of finding two different 
configurations in the same minimum and it is given by 
\be
a_{N}=\sum_{\al }w_{\al}^{2}.
\ee

In the one step replica symmetry breaking hypothesis two  
configurations that are not in the same minimum are expected to be orthogonal,
i.e. at the maximum possible distance $d_{1}$. 
This happens with the probability:
\be
b_{N}=\sum_{\al \neq \ga} w_{\al}w_{\ga}.
\ee

The functions $a_{N}$ and $b_{N}$ are naturally depending on the temperature.
In the mean field framework one finds that the weight of the $\delta$-function
in $d_{1}$, $b_{N}$, that is zero for $T \geq T_{K}$, increases 
continuously when lowering the temperature below the transition point.

The probability distribution $P_{N}(d)$ should therefore be an appropriate
observable for looking at the transition from the high-$T$ region 
(where $P_{N}(d)$ is Gaussian-like for a finite system) to the glassy phase in which
it is expected to show a non trivial behaviour, strongly depending on $N$.

\subsection{The approach to equilibrium}
\noindent
For simplicity we will point the attention on the relaxation of the energy 
density when the system is quenched abruptly (at the time $t=0$) from a 
random initial 
configuration (i.e. infinite cooling rate) to the final temperature. 

In the mean field picture \cite{cuku} one finds two different 
relaxation behaviours, depending on the final temperature value:
\begin{itemize}
\item At high temperatures the energy reaches exponentially its equilibrium
value, i.e. $e(t)=e_{eq}+ c \: \exp(-t/ \tau)$.
The relaxation time $\tau$ 
increases when lowering the temperature and it diverges for 
$T \rightarrow T_{D}$.

\item Below the dynamical transition point $T_{D}$ the energy behaves 
linearly as a function of $t^{-\al}$ \cite{FRAMAPA}, i.e.
$e(t)=e_{D}+c \: t^{-\al}$, 
the exponent 
$\al$ being weakly depending on $T$. Here the system is evolving towards
some metastable states (with infinite life time) and correspondingly the
asymptotic energy value $e_{D}$ is higher than that of the equilibrium.
\end{itemize}

The infinite life time of metastable states is just an artefact of the
mean field approximation. In a real system we expect the approach
to the equilibrium for $T < T_{D}$ consisting of two steps:
\begin{itemize}
\item The convergence to some metastable states with a mean field like 
behaviour.
\item The slow decay of metastable states due to activated processes.
In this second step the system reaches the true equilibrium state that
will be still the replica-symmetric one for $T \geq T_{K}$.
\end{itemize}

This means that by looking at $e(t)$ it is in principle possible to find 
numerical evidence for the reminiscence in real glasses of the mean field 
dynamical transition. Here $T_{D}$ is expected to mark the onset of the
two steps relaxation.

\section{The model}

\noindent
We study a binary mixture of soft spheres, half of the particles  
being of type $A$ with radius $\sigma_A$ and half of type $B$ with radius 
$\sigma_B$. The Hamiltonian is:
\begin{equation}
{\cal H}_{pbc} = \sum_{i < k=1}^{N} \left ( \frac{ \sigma(i) + \sigma (k)}
{ | \mbox{\boldmath $r$}_i - \mbox{\boldmath $r$}_k | } \right )^{12}.
\end{equation}
This model has been carefully studied in the past \cite{HANSEN1}-\cite{softl}.
It is known that the choice $\sigma_B / \sigma_A = 1.2$ strongly inhibits 
crystallization. We also follow the convention of considering particles with 
average diameter 1 by setting:
\begin{equation}
{\sigma_A^3+2(\sigma_A+\sigma_B)^3+\sigma_B^3}=\frac{1}{2}.
\end{equation} 
Thermodynamic quantities only depend on $\Gamma \equiv \rho / T^{1/4}$, where 
$T=1 / \beta$ is the temperature and we have taken density $\rho=1$  
($\Gamma \equiv \beta^{1/4}$). The $N$ particles move in a $3d$ cube of size 
$L=N^{(1/3)}$. The glass transition is known \cite{HANSEN2} to happen 
around $\Gamma_c=1.45$.

In order to obtain numerical results on the equilibrium properties
 comparable with the mean 
field theoretical picture we attempt to measure the equilibrium probability 
distribution $P(d)$ of the distance $d$ between states. Following the usual 
strategy in spin glass simulations we introduce two evolving 
contemporaneously and independently replicas of the
system. $P(d)$ is then given by:
\be
P(d) \equiv \lan \delta ( d- {\cal D} ) \ran,
\ee
${\cal D}$ being the appropriately defined distance between the configurations 
of the two replicas.

Labeling by $\{ \mbox{\boldmath $r$}_i \}$, $\{ \mbox{\boldmath $s$}_i \}$ 
the positions of the $N$ particles in the two replicas, a natural 
definition of ${\cal D}$ is the Euclidean one, minimized over permutations 
$\pi$, rotations and reflections {\small \tt R} and, 
when using periodic boundary conditions, 
translations {\small \tt T}:
\begin{equation}
{\cal D}^2  \equiv  \frac{1}{N} \min_{\pi,\mbox{\small \tt R,T}} \left 
( \sum_{type \: A} 
 ( \mbox{\boldmath $r$}_i - \mbox{\boldmath $s$}_{\pi(i)} )^2 
+ \sum_{type \: B}  ( \mbox{\boldmath $r$}_i - 
 \mbox{\boldmath $s$}_{\pi(i)} )^2 \right ).
\end{equation}
An analogous definition of distance between configurations of particles
has been considered in \cite{heuer} and in \cite{apr}, in studies of 
potential energy minima for Lennard-Jones systems.

We minimize over permutations by an approximate procedure. For each 
particle $i$ of one configuration we take $\pi(i)$ being the nearest one of 
the 
other. We expect this to be a reasonable approximation in the considered 
temperature range (from $\Gamma$=1 to $\Gamma$=2) since the probability for 
two particles of the same system of being at a distance lower than 
$2 \sigma \simeq 1$ is very close to zero at not too high temperatures 
(the radial density-density correlation function in a simple liquid shows the 
well known behaviour $g(r) \simeq 0$ for $r < 2 \sigma$).

Minimization over rotations and reflections is easily performed since 
in the case of a 
$3d$ cube {\small \tt R} results a discrete group that includes 48 symmetry 
operations, corresponding to the $2^3$ reflections and to the $3!$ 
permutations of axes. On the other hand, minimization over the continuous 
group of translations is 
a hard task and we prefer to avoid it by not using periodic boundary 
conditions. 

To measure $P(d)$ we have considered a slightly modified model in which 
particles are definitely confined in the cubic box of size $L$ by a soft 
walls-repulsive potential term:
\begin{equation}
{\cal H}_{sw}= {\cal H}_{pbc} + c_1 \sum_{i=1}^{N} \sum_{\mu=1}^{3}
\left ( \frac{1}{(c_2 + {r_i}^{\mu})^{10}} +\frac{1}{(L + c_2- 
{r_i}^{\mu})^{10}}  \right ).
\label{sw}
\end{equation}
We have chosen the values $c_1= \pi / 5$ and $c_2 = 0.6$ that give a behaviour 
of the energy as a function of $\Gamma$ quite near to that of the periodic 
boundary conditions case. 

\section{Numerical results on the dynamics}

\subsection{Algorithms}

\noindent
Both for finding the best numerical approach to the simulations of structural
glasses at the equilibrium and for studying the dynamical properties of the
model we have implemented different algorithms, considering 
stochastic and deterministic dynamics:
\begin{itemize}
\item Monte Carlo ($MC$). We start from a random configuration and we 
quench abruptly the system by putting it at the final temperature (i.e. 
infinite cooling rate). During one step  each particle is suggested to move 
of a random quantity and the maximum shift permitted is chosen in order to 
obtain an acceptance ratio near 0.5. 
\item Molecular Dynamics ($MD$). When using 
$MD$ to start with a completely random spatial configuration may cause 
difficulties. Here the initial spatial configuration is obtained from a 
random one by 60 Monte Carlo steps at the final temperature. At the beginning 
and again each 100 steps we extract momenta $\{ \mbox{\boldmath $p$}_i \}$ 
according to the Boltzmann distribution (we have taken the masses of particles 
$m_A=m_B=1$) and we impose the condition $\sum_{i=1}^{N} 
{\mbox{\boldmath $p$}_i}^2=(3N- f_{MD})T$, where $f_{MD}$ is the number of 
frozen degrees of freedom. $f_{MD}=3$ when using periodic boundary conditions 
since the three components of the total linear momentum 
$\mbox{\boldmath $P$}$ are conserved (we have taken 
$\mbox{\boldmath $P$}=\mbox{\boldmath 0}$), otherwise $f_{MD}=0$.  
We use the velocity-Verlet algorithm \cite{md} with $\delta t =1/250$ 
(a value that we have checked to be reasonable in our case, e.g. 
the total energy results perfectly constant).
\item Isothermal Molecular Dynamics ($IMD$). The system evolves 
according to the Gaussian isokinetic equations of motion \cite{md}:
\begin{equation}
\dot{\mbox{\boldmath $r$}}_i = \mbox{\boldmath $p$}_i  \hspace{1cm} 
\dot{\mbox{\boldmath $p$}}_i = \mbox{\boldmath $F$}_i - 
\lambda \mbox{\boldmath $p$}_i
\hspace{1cm} \lambda=\frac{\sum_{i=1}^{N} \mbox{\boldmath $F$}_i \cdot 
\mbox{\boldmath $p$}_i}{\sum_i 
{\mbox{\boldmath $p$}_i}^2}.
\end{equation}
At the beginning we extract momenta and we fix the constraint $\sum_{i=1}^{N} 
{\mbox{\boldmath $p$}_i}^2=(3N-f_{IMD})T$, where $f_{IMD}=f_{MD}+1$. 
Also in this case we start from a spatial configuration obtained from a random 
one by 60 Monte Carlo steps at the final temperature. We implement the 
algorithm by using a variant of the leap-frog scheme \cite{md} again with 
$\delta t =1/250$.
\item Parallel Tempering ($PT$). 
Algorithms in which temperature is allowed to become a dynamical variable
\cite{st} are very effective for thermalizing systems with a complex 
free energy landscape. In this recently introduced \cite{pt} method a set of 
$n$ different $\beta$ values $ \beta_1 < \cdots < \beta_k  \cdots \beta_n$ is 
chosen $a \: priori$ and $n$ replicas of the system evolve contemporaneously. 
The extended Hamiltonian ${\cal H}_{PT}= \sum_{a=1}^{n} \beta(a) \: 
{\cal H} [ C_a ]$ is defined and exchanges of temperatures between replicas 
$$( \beta(a_1)=\beta_{k_1}, \beta(a_2)= \beta_{k_2} ) 
\rightarrow ( \beta(a_1)= \beta_{k_2}, \beta(a_2)=\beta_{k_1} )$$ are allowed 
with probability $p= \min [1, \exp (- \Delta {\cal H}_{PT}) ] $.
The whole process is itself a Markov chain and, when equilibrium has been 
reached, each replica moves between different temperatures of the set 
remaining at the equilibrium.
We start by extracting independently the $n$ random initial configurations and 
we quench each replica at a different one of the chosen temperatures. In a 
$PT$ step, sequentially for $a=1 \dots n$: 
\begin{itemize}
\item
Replica $a$ makes one $MC$ step at its temperature 
$\beta (a)= \beta_k$. 
\item
A random number $j= \pm 1$ with equal probability is extracted.  
\item
For $ 1 \leq k+j \leq n$ the exchange of temperatures between 
replica $a$ and replica $b$, where  $\beta(b)= \beta_{k+j}$, is suggested and 
possibly accepted. 
\end{itemize}
The set of $\beta$ values should be chosen carefully. We find that for the 
numbers of particles considered (up to $N$=36) $PT$ works well down to 
$\Gamma$=2 with $n$=13 different temperatures ($\Gamma$=1, 1.05 
\dots 1.2, 1.3 $\dots$ 2).
\item We have implemented and tested a combined technique of 
Parallel Tempering and Isothermal Molecular Dynamics ($IMDPT$). 
The $n$ initial spatial configurations are obtained from random ones by 60 
$MC$ steps at $\Gamma$=1. 
After extracting momenta we impose the constraints $\sum_{i=1}^{N} 
({\mbox{\boldmath $p$}_i}^a)^2=(3N-f_{IMD})T_a \: \: a=1 \dots n$ (we have 
taken the same set of temperatures that in $PT$). Replicas evolve accordingly 
to $IMD$ and each 10 $IMD$ steps the exchanges of temperatures happen with 
probability $p= \min [1, \exp(- \Delta {\cal H}_{PT})]$, where in 
${\cal H}_{PT}$ 
only potential energies appear. Here to change temperature $T_{old} 
\rightarrow T_{new}$ means to transform momenta as $\mbox{\boldmath $p$}_i 
\rightarrow (T_{new} / T_{old})^{1/2} \mbox{\boldmath $p$}_i, \: \: 
i=1 \dots N$.
\item Finally we have performed $MC$ simulations of large systems 
(2000 and 8000 particles) in the periodic boundary conditions case, 
by putting a cut-off on the soft spheres' potential:
\be
V_{ik}= \left \{
\begin{array}{ccc}
 \left ( \frac{ \sigma(i) + \sigma (k)}
{ | \mbox{\boldmath $r$}_i - \mbox{\boldmath $r$}_k | } \right )^{12}
& \hspace{.3cm} & \mbox{for 
$| \mbox{\boldmath $r$}_i - \mbox{\boldmath $r$}_k | < R$} \\
\hspace{.3cm} &\hspace{.3cm} & \hspace{.3cm}\\
 \left ( \frac{ \sigma(i) + \sigma (k)}
{ R } \right )^{12} = V_{ik}^{0}
& \hspace{.3cm} & \mbox{otherwise} \\
\end{array}
\right.
\ee
The algorithm is then implemented in such a way that for each particle 
the map of the ones which are at distance lower than $R+ 2 \delta$ is 
recorded and updated during the run ($\delta$ being the maximum shift 
permitted to a particle in one $MC$ step). We have chosen $R=1.7$ that
means a practically negligible $V_{ik}^{0} \sim O(10^{-3})$ ($V_{ik} \sim 1$
for $| \mbox{\boldmath $r$}_i - \mbox{\boldmath $r$}_k | \sim 2 \sigma$). 
\end{itemize}

\subsection{On the energy relaxation behaviour}

\noindent
We look at the (potential) energy density relaxation when the system is 
quenched abruptly from a random initial configuration to the final (low) 
temperature. We compare results obtained by different numerical methods for
$N$=34 particles and we extend the analysis to large systems ($N$=2000 and 
$N$=8000) in the $MC$ case. Here periodic boundary conditions are used. 

In [Figs. 1-3] we present data obtained by $MC$, $MD$ and $IMD$, respectively.
Our results show no evident difference between stochastic and deterministic 
dynamics. Not only in the $MC$ case, as already observed in previous 
simulations on the same model \cite{soft2}, but also when using Molecular 
Dynamics techniques the energy density relaxation is well compatible on a 
large time window with a linear behaviour in function of $t^{-\alpha}$. 
We get a non small exponent $\alpha \sim 0.8$, weakly $\Gamma$-dependent
in the considered range (the dependence seems slightly more 
pronounced in the $IMD$ case). 

\begin{figure}[htpb]
\vspace{-2.2cm}
\begin{center}
\leavevmode
\centerline{\epsfig{figure=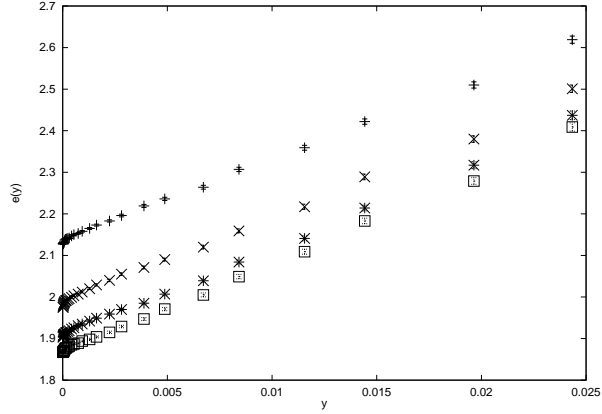,angle=270,width=8cm}} 
\end{center}
\vspace{-.7cm}
\caption{$MC$ data on $e$ as a function of $y=t^{(-0.8)}$ at $\Gamma$=1.4(+),
$1.6 (\times)$, $1.8 (\ast)$ and $2.0 (\Box)$. Here 150 different initial 
conditions are considered.}
\label{fig:1}
\end{figure}

\begin{figure}[htbp]
\vspace{-1.8cm}
\begin{center}
\leavevmode
\centerline{\epsfig{figure=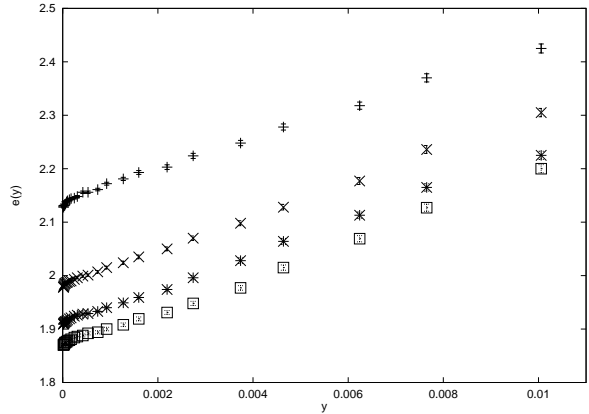,angle=270,width=8cm}} 
\end{center}
\vspace{-.7cm}
\caption{$MD$ data on $e$ as a function of $y=t^{(-0.8)}$ at $\Gamma$=1.4(+), 
$1.6 (\times)$, $1.8 (\ast)$ and $2.0 (\Box)$. Here 100 different initial 
conditions are considered. When using Molecular Dynamics techniques the need 
of a non completely random initial configuration (i. e. the first 60 steps are
$MC$ steps) induces some short time effects and the linear behaviour takes 
place at a slightly larger time than in the $MC$ case.}
\label{fig:2}
\end{figure}

\begin{figure}[htbp]
\vspace{-1.8cm}
\leavevmode
\centerline{\epsfig{figure=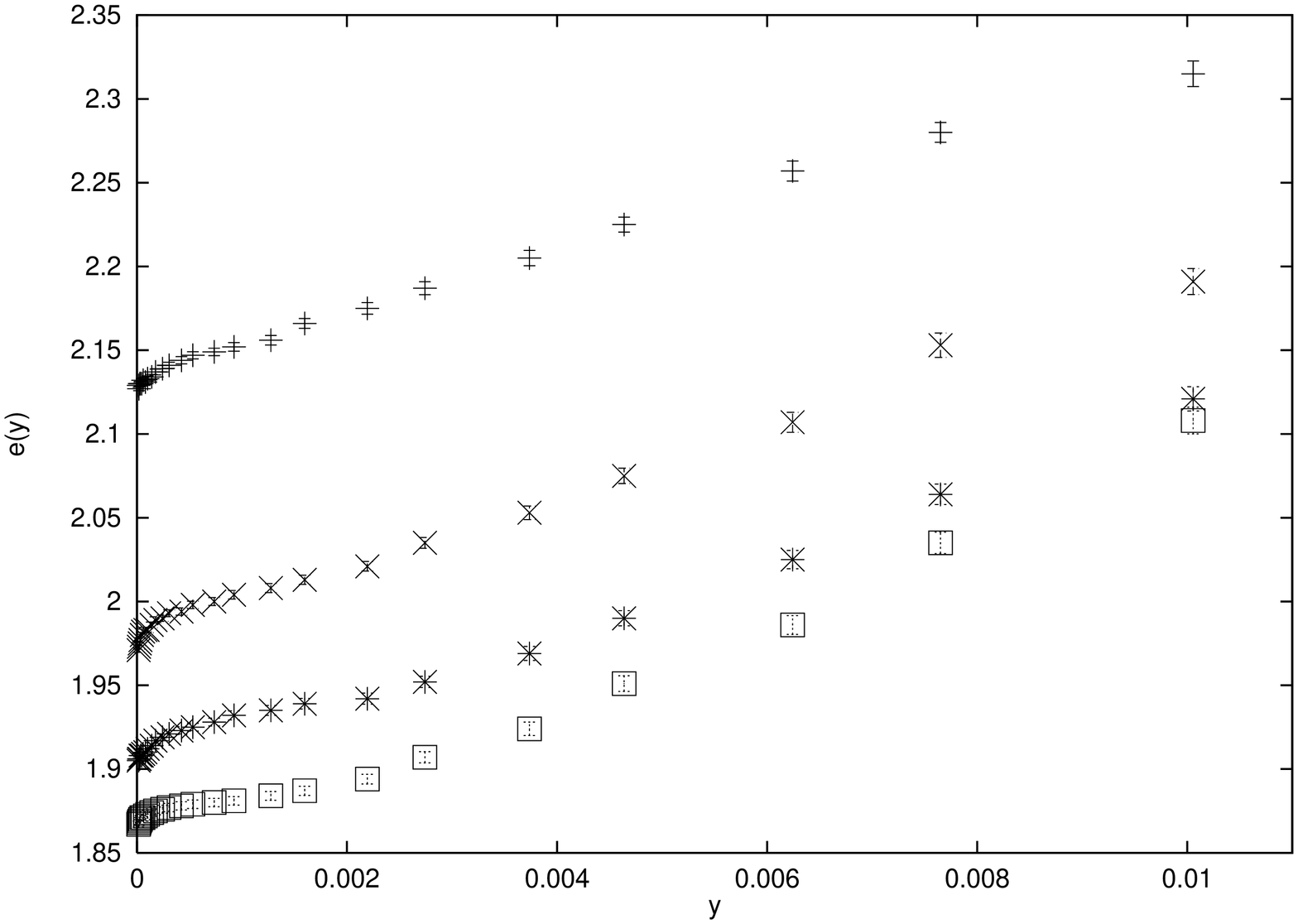,angle=270,width=8cm}}
\vspace{-.7cm} 
\caption{$IMD$ data on $e$ as a function of $y=t^{(-0.8)}$ at 
$\Gamma$=1.4(+), $1.6 (\times)$, $1.8 (\ast)$ and $2.0 (\Box)$. 
Here 100 different initial conditions are considered. The dependence
of $\alpha$ on $\Gamma$ seems slightly more pronounced in this case, our 
best estimates ranging from $\alpha \sim 0.8$ at $\Gamma$=1.4 to $\alpha 
\sim 1.1$ at $\Gamma$=2.0.}
\label{fig:3}
\end{figure}

Moreover, the energy values at a fixed temperature that one can estimate by 
asymptotically extrapolating the linear behaviours obtained by different 
methods, are near to be the same. On the other hand we are going to show that 
these asymptotic values (apart from the case of $\Gamma$=1.4) are not the real 
equilibrium energy values of the system.

We present in [Figs. 4-5] $PT$ and $IMDPT$ data. There is a reminiscence of 
the linear behaviour but here the system moves between the high and the low 
temperatures of the set, this preventing it from being trapped. At the end
of the (quite large) time window, equilibrium is near to be reached and the
energy values are definitely smaller than the ones estimable by extrapolating 
simple Monte Carlo or Molecular Dynamics data. The difference is already 
detectable at $\Gamma$=1.6 and it becomes more pronounced at lower 
temperatures. To further outline this result we plot in [Fig. 6] data obtained 
by various methods at $\Gamma$=2.0.

\begin{figure}[htbp]
\begin{center}
\leavevmode
\centerline{\epsfig{figure=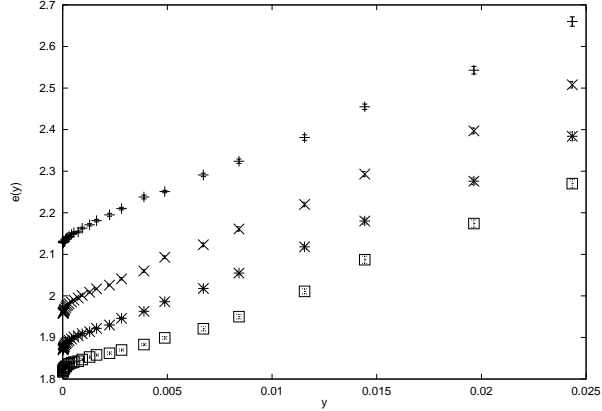,angle=270,width=8cm}} 
\end{center}
\caption{$PT$ data on $e$ as a function of $y=t^{(-0.8)}$ at 
$\Gamma$=1.4(+), $1.6 (\times)$, $1.8 (\ast)$ and $2.0 (\Box)$. Here 24
different initial conditions are considered. Note that both in this case and 
in the $IMDPT$ one the obtained behaviour depends on the entire set of 
temperatures.}
\label{fig:4}
\end{figure}

\begin{figure}[htbp]
\begin{center}
\leavevmode
\centerline{\epsfig{figure=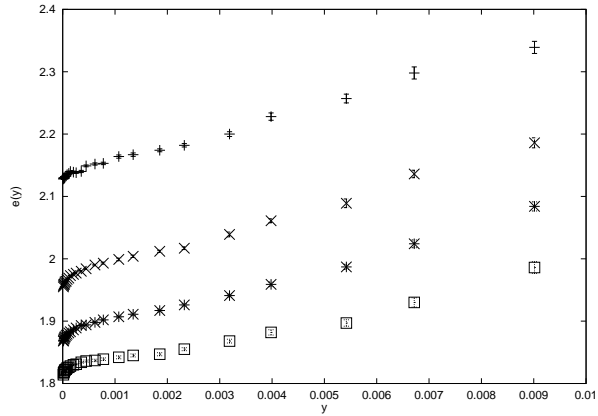,angle=270,width=8cm}} 
\end{center}
\caption{$IMDPT$ data on $e$ as a function of $t^{(-0.8)}$ at 
$\Gamma$=1.4(+), $1.6 (\times)$, $1.8 (\ast)$ and $2.0 (\Box)$. Here 24
different initial conditions are considered. Data are near to be compatible 
within the errors with the $PT$ ones.}
\label{fig:5}
\end{figure}

\begin{figure}[htbp]
\begin{center}
\leavevmode
\centerline{\epsfig{figure=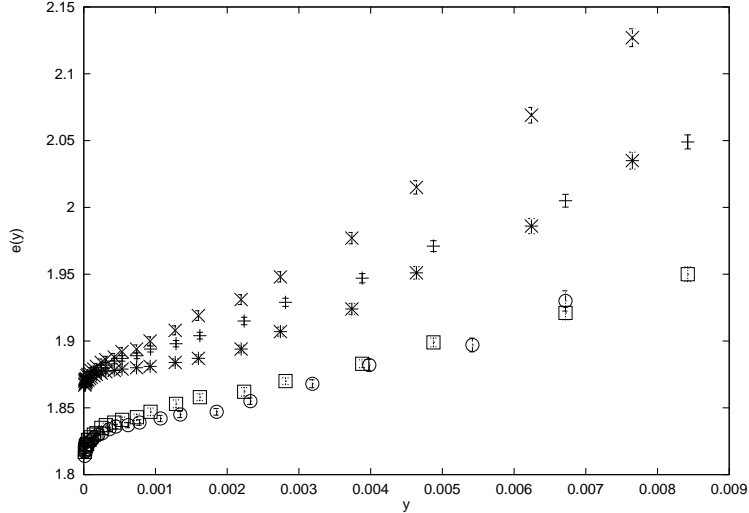,angle=270,width=10cm}} 
\end{center}
\caption{Data on $e$ as a function of $t^{(-0.8)}$ at $\Gamma$=2.0 obtained by
$MC (+)$, $MD (\times)$, $IMD (\ast)$, $PT (\Box)$ and $IMDPT (\circ)$.}
\label{fig:6}
\end{figure}

We have found numerical evidence for the energy relaxation  
at low temperatures
consisting of two processes that happen on remarkably well separated 
time scales. Both in the $MC$ and in the Molecular Dynamics data 
the first step is clearly
observable, corresponding to the convergence to some metastable 
states 
with a mean field like behaviour (note that $\alpha$ is weakly depending on 
the temperature on a large range). The fact that the extrapolated energy 
values are
not the real equilibrium values gives evidence for the presence of a 
second step, the slow decay of metastable states dominated by activated 
processes.

These results are inadequate for understanding whether the curvature of
low temperature data on
$e(y)$ at very large times (small $y=t^{-\alpha}$) is related to the 
quite small $N$ value we are considering or it actually represents the 
beginning of the second step (possibly with an intermediate $plateau$). 
In order to clarify this point we consider definitely larger systems.

We plot in [Fig. 7] $e(y)$ as obtained by $MC$ simulations for $N$=8000 
at $\Gamma$=1.4, 1.5, 1.8 and 3.0. At $\Gamma$=1.8 we present data also
for $N$=2000 that result almost indistinguishable from the $N$=8000 ones 
and remarkably similar to $MC$ data for the well smaller $N$=34 system.
This means that we are actually looking both at the first relaxation
step and at the beginning of the second one. 

The different behaviours observed when varying the temperature agree
well with the previously discussed theoretical picture:
\begin{itemize}
\item At $\Gamma$=1.4 no two steps behaviour is observable and the
energy reaches the equilibrium value in the considered time window.
The system appears to be still above $T_{D}$ but quite near to it
(the relaxation time $\tau$ seems to be very large, the expected 
exponential decay being not distinguishable from a linear behaviour 
in function of $y$). 
\item For $\Gamma \geq 1.5$ the system is definitely below $T_{D}$.
The energy relaxes linearly as a function of $y$ to a 
value $e_{D}$ higher than that of equilibrium, the slow decay of 
metastable states happening on a well
larger time scale.
\item The beginning of the second step is clearly observable at
$\Gamma$=1.5 and still quite evident at $\Gamma$=1.8 but it
disappears at $\Gamma$=3.0. Here data behave linearly on the entire
time window. This seems reasonable since the time
scale that control the second step (i.e. the onset of activated processes) 
is expected to increase when lowering the temperature.
\end{itemize}

\begin{figure}[htbp]
\begin{center}
\leavevmode
\centerline{\epsfig{figure=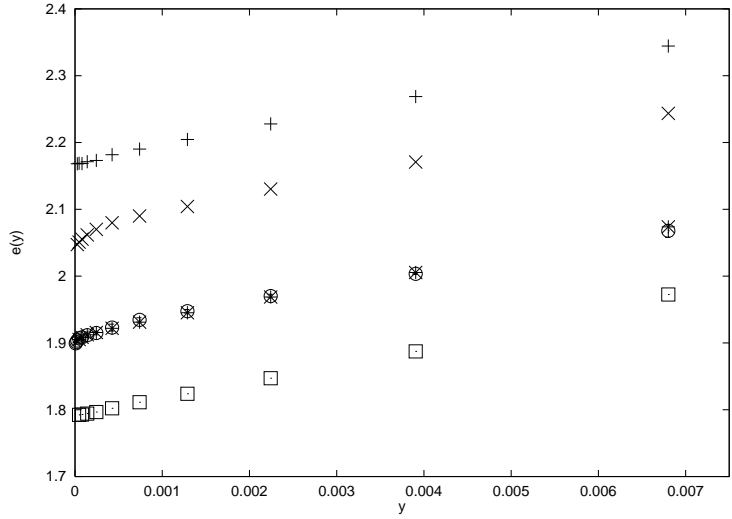,angle=270,width=10cm}} 
\end{center}
\caption{$MC$ data on $e$ as a function of $t^{(-0.8)}$ respectively 
for $N$=8000 (one initial condition) at $\Gamma$=1.4 (+), $1.5 (\times)$, 
$1.8 (\ast)$ and $3.0 (\Box)$ and for $N$=2000 (two different 
initial conditions) at $\Gamma$=1.8 $(\circ)$.}
\label{fig:7}
\end{figure}

\begin{figure}[htbp]
\begin{center}
\leavevmode
\centerline{\epsfig{figure=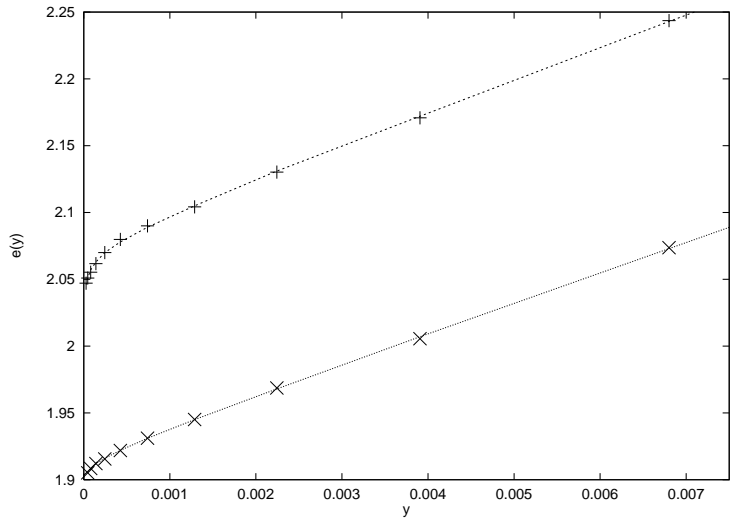,angle=270,width=10cm}} 
\end{center}
\caption{$MC$ data on $e$ as a function of $t^{(-0.8)}$ for
$N$=8000 at $\Gamma$=1.5 (+) and $1.8 (\times)$. The lines are 
the corresponding
best fits to $e(t)=a t^{(-0.8)} + b t^{\gamma} +c$ obtained by the plotted 
points.}
\label{fig:8}
\end{figure}

To fit the two steps 
behaviour is a hard task. In [Fig. 8] we show
our best results at $\Gamma$=1.5 and 1.8 ($N$=8000), obtained by
fixing $\alpha=-0.8$ and considering $e(t)=a t^{(-0.8)} + b t^{\gamma} +c$
It is interesting to note that in both cases we get a positive $\gamma$
value of the same order, $\gamma \sim +0.1$. It seems reasonable that the
second step corresponds to a streched exponential decay, i.e.
 $e(t)=a_{1} (t / \tau_{1})^{- \alpha } +
a_{2} \exp( - C (t / \tau_{2} )^{+ | \gamma | }) +e_{eq}
\simeq a t^{(-0.8)} - |b| t^{+ |\gamma|} 
+ e_{plateau}$ for $t \ll \tau_{2}$. However  
we stress that this is a purely indicative result
since quite different $\gamma$ values (possibly of the opposite sign) 
are obtainable by slightly varying $\alpha$ or the time window.

To conclude this section, we note that $IMDPT$ works well but seems to be 
not more effective than usual $PT$. We have done some 
dynamical simulations with different methods also in the case of 
Hamiltonian (\ref{sw}). We do not show the results that are qualitatively 
similar to those obtained by using periodic boundary conditions, 
the only difference being 
that here the onset of the second step becomes evident at shorter times.

\section{Numerical results on the statics}

\subsection{Simulations at the equilibrium}
\noindent
In order to get the equilibrium behaviour of $P(d)$ we use Parallel Tempering,
simulating contemporaneously two independent sets of replicas. Thermalization 
is a hard task and we limit our analysis to quite small numbers of particles 
ranging from $N$=28 to $N$=36. For each $N$ value we perform an extensive 
simulation of $2^{22} \div 2^{24}$ $PT$ steps (i. e. up to more than 16 
millions of $MC$ steps for each one of the 26 replicas). We measure all the 
quantities we are interested in during the last 3/4 of the run. 

Thermalization is checked in different ways:
\begin{itemize}
\item We divide the last part of the run, in which statistics is 
collected, into 16 equal intervals and we look for possible shifts of 
the corresponding mean values (particularly we do not find evident changes 
in the behaviour of $P(d)$). 
\item We check that each replica moves more than once from an extrema of 
the temperature range to the other and back in the last part of the run. 
\item We valuate the specific heath $c$ both using $ c= 
\partial < e > / \partial T$ and using $T^2 \: c = < e^2 > - < e >^2$, 
checking compatibility of results.
\end{itemize}

Errors are estimated from the mean values obtained in each one of the 16 
intervals.

Despite our efforts it is not possible to exclude the presence of 
lower minima in the free energy landscape that are not accessible to the 
system on the considered time scale. To clarify this point it is interesting to
look at the particular value $N$=32=$4^3/2$. Here particles are allowed to 
fill all the sites of a $fcc$ structure. Once this crystalline 
equilibrium state has been reached (in about $2^{24} \div 2^{25}$ $PT$ 
steps) the system rests definitely trapped in it, the crystalline 
configuration corresponding to a ``golf-hole'' shaped minimum in the 
free energy landscape. The energy density results a discontinuous function of 
$\Gamma$ as expected for a first order (liquid-crystal) transition.

It seems unlikely to us but particles could be able to arrange themselves 
in some sort of crystalline configuration that we have not detected, also for 
$N \neq 32$. On the other hand we would like to look at the glass 
equilibrium states, apart from the possible presence of lower crystalline 
minima. We find therefore reasonable to assume that on the time scale 
studied the system extensively explores the entire phase-space region we are 
interested in.

\subsection{On the behaviour of $P(d)$}

Data on the energy density $e(\Gamma)$ for different $N$ values are plotted
in [Fig. 9]. We note that the equilibrium energy density is a not regular 
function of the number of particles since it is definitely smaller for $N$=34 
than for $N$=36. The non trivial $N$-dependence of the model at low 
temperatures is outlined when looking at the specific heath $c(\Gamma)$ 
[Fig. 10]. 

\begin{figure}[htbp]
\begin{center}
\leavevmode
\centerline{\epsfig{figure=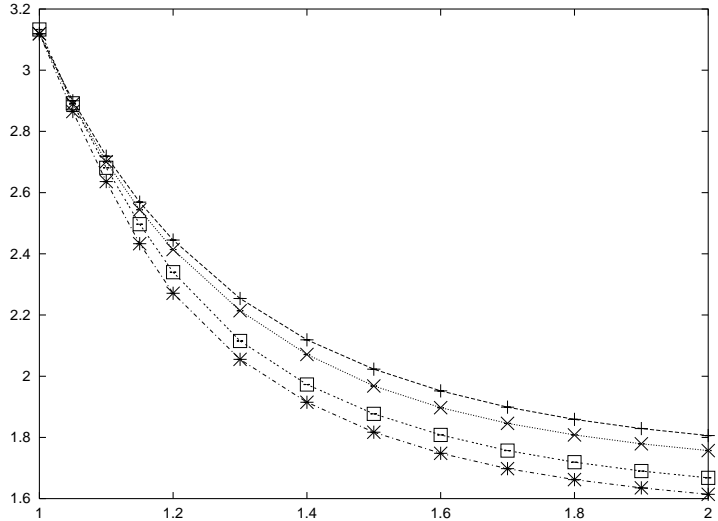,angle=270,width=10cm}} 
\end{center}
\caption{Data on the energy density $e$ as a function of $\Gamma$ for 
$N$=28(+), $30 (\times)$, $34 (\ast)$ and $36 (\Box)$. Lines are only to 
join neighbouring points.}
\label{fig:9}
\end{figure}
\nopagebreak
\begin{figure}[htbp]
\begin{center}
\leavevmode
\centerline{\epsfig{figure=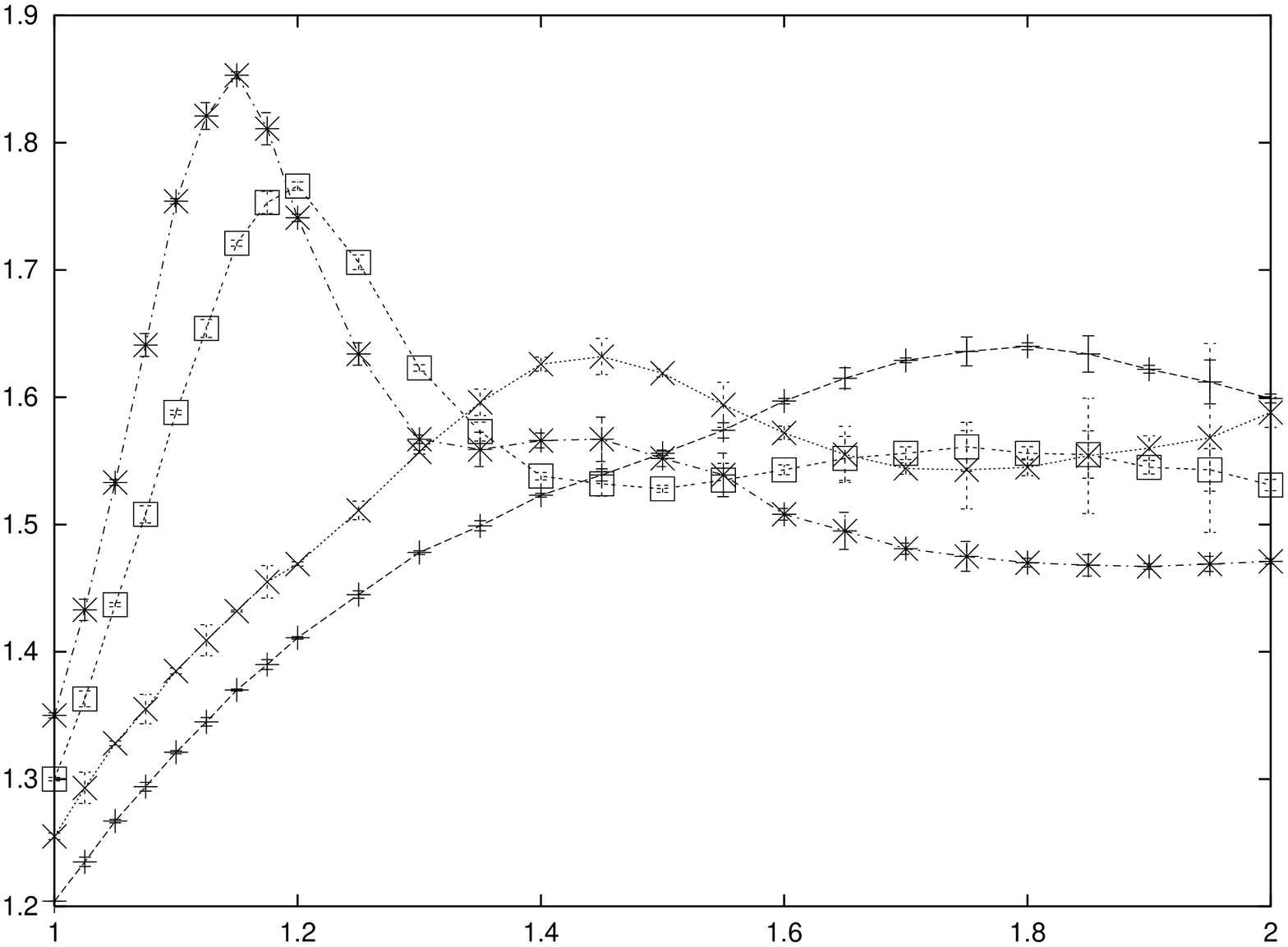,angle=270,width=10cm}} 
\caption{Data on the specific heath $c$ as a function of $\Gamma$ for 
$N$=28(+), $30 (\times)$, $34 (\ast)$ and $36 (\Box)$. Lines are only to 
join neighbouring points.}
\end{center}
\label{fig:10}
\end{figure}

It should be stressed that for each given $N$ value the energy appears a 
continuous function of $\Gamma$ and correspondingly the behaviour of 
$c(\Gamma)$ looks quite smooth, this giving evidence for the physics 
phenomenon we are studying being not a crystallization process. Moreover, we 
never observed the system definitely trapped in a configuration as it 
happens in the previously discussed $N$=32 case.

We plot in [Fig. 11] data on $P(d)$ for different $N$ values at the highest 
temperature considered ($\Gamma$=1). The shape is the Gaussian-like one 
characteristic of the liquid phase and it changes very little when varying the
number of particles.

\begin{figure}[htbp]
\begin{center}
\leavevmode
\centerline{\epsfig{figure=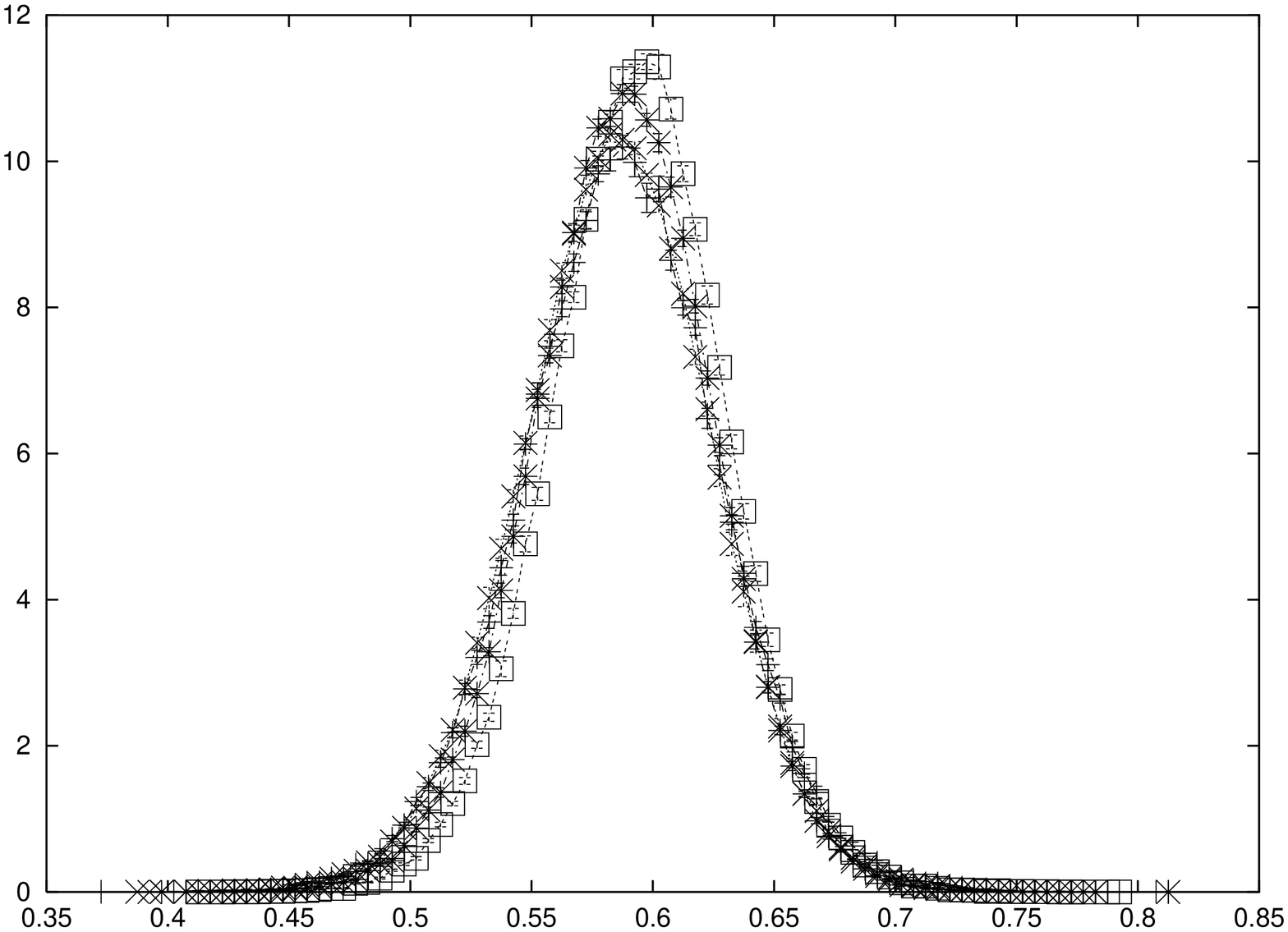,angle=270,width=10cm}} 
\caption{Data on $P(d)$ at $\Gamma$=1 for $N$=28(+), $30 (\times)$, $34 
(\ast)$ and $36 (\Box)$.}
\end{center}
\label{fig:11}
\end{figure}

Finally we present in [Figs. 12-15] the low temperature data on $P(d)$
for $N$=28, 30, 34 and 36, respectively. The qualitative features are the
same for different number of particles. The behaviour changes in a remarkably 
short $\Gamma$ range around $\Gamma$ = $\Gamma^{*}$, and $P(d)$ 
looks highly non 
trivial at higher $\Gamma$ values. The physical meaning of the peaks shown by 
the equilibrium probability distribution of distance between states 
in the low temperature region comes from the mean field theoretical picture. 
In this region the data are well consistent with the scenario that in the
glassy phase a small number of valleys in the free energy landscape give 
the dominant contribution to the partition function. 

\begin{figure}[htbp]
\begin{center}
\leavevmode
\centerline{\epsfig{figure=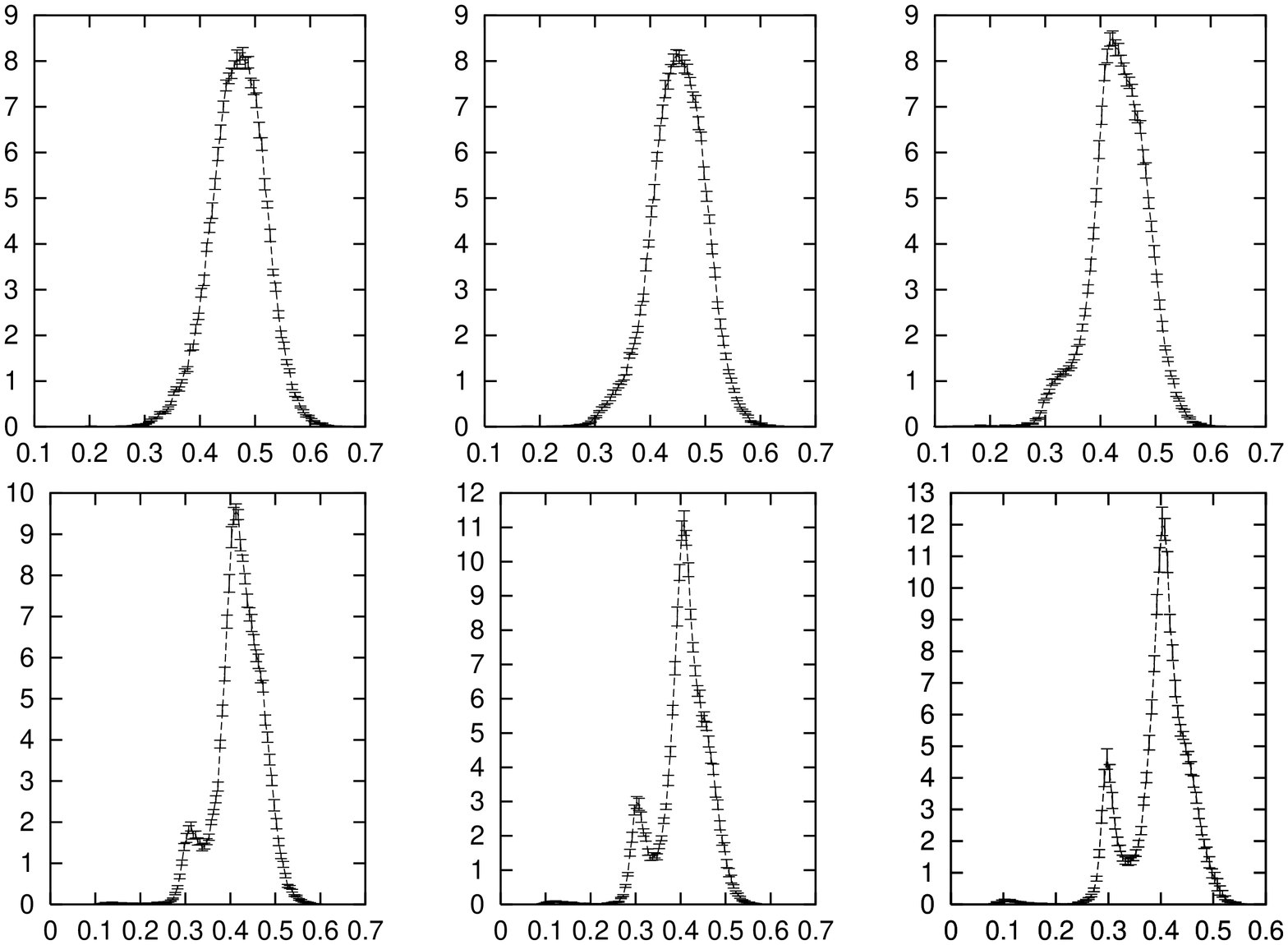,angle=270,width=10cm}} 
\caption{Data on $P(d)$ for $N$=28. From left to right and top to bottom 
$\Gamma$=1.5, 1.6, 1.7, 1.8, 1.9 and 2. Plotted data have been measured
during a $2^{24}PT$ steps run but we stress that they are perfectly
compatible with those obtained by a previous $2^{22}$ step run.}
\end{center}
\label{fig:12}
\end{figure}
\nopagebreak
\begin{figure}[htbp]
\begin{center}
\leavevmode
\centerline{\epsfig{figure=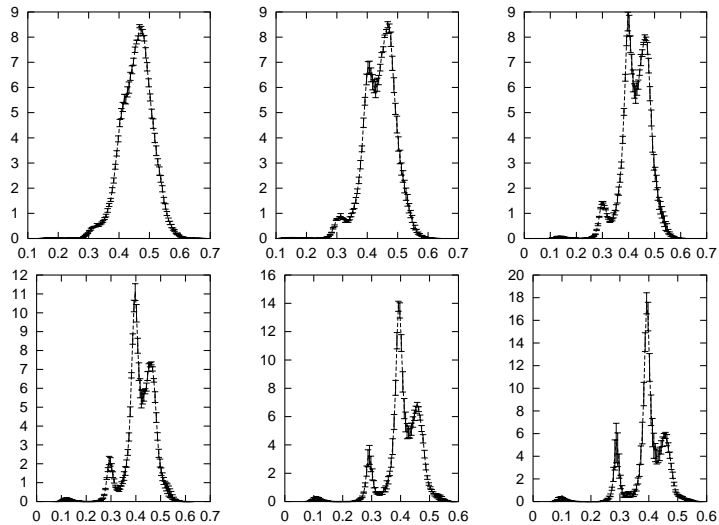,angle=270,width=10cm}} 
\caption{Data on $P(d)$ for $N$=30. From left to right and top to bottom 
$\Gamma$=1.5, 1.6, 1.7, 1.8, 1.9 and 2 (i. e. the same value that in 
previous [Fig. 10]). Here data have been measured during a $2^{22} PT$ 
step run.}
\end{center}
\label{fig:13}
\end{figure}

\begin{figure}[htbp]
\begin{center}
\leavevmode
\centerline{\epsfig{figure=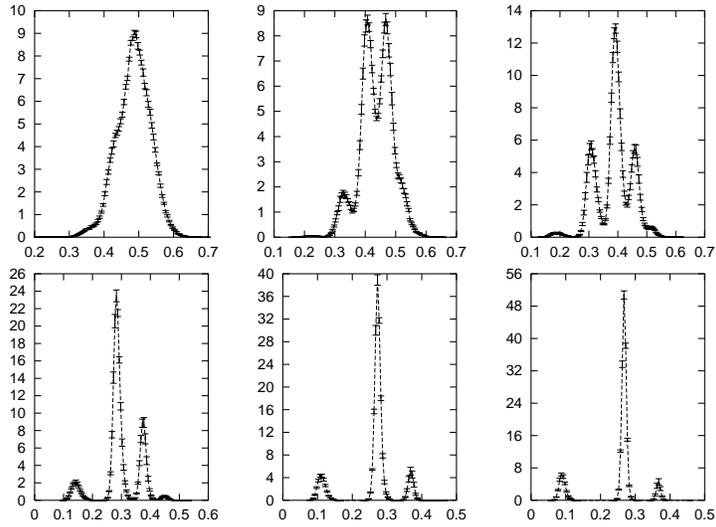,angle=270,width=10cm}} 
\caption{Data on $P(d)$ for $N$=34. From left to right and top to bottom 
$\Gamma$=1.2, 1.3, 1.4, 1.6, 1.8 and 2. Here and in the next case 
of $N$=36 data have been measured during a $2^{24} PT$ step run.}
\end{center}
\label{fig:14}
\end{figure}
\nopagebreak
\begin{figure}[htbp]
\begin{center}
\leavevmode
\centerline{\epsfig{figure=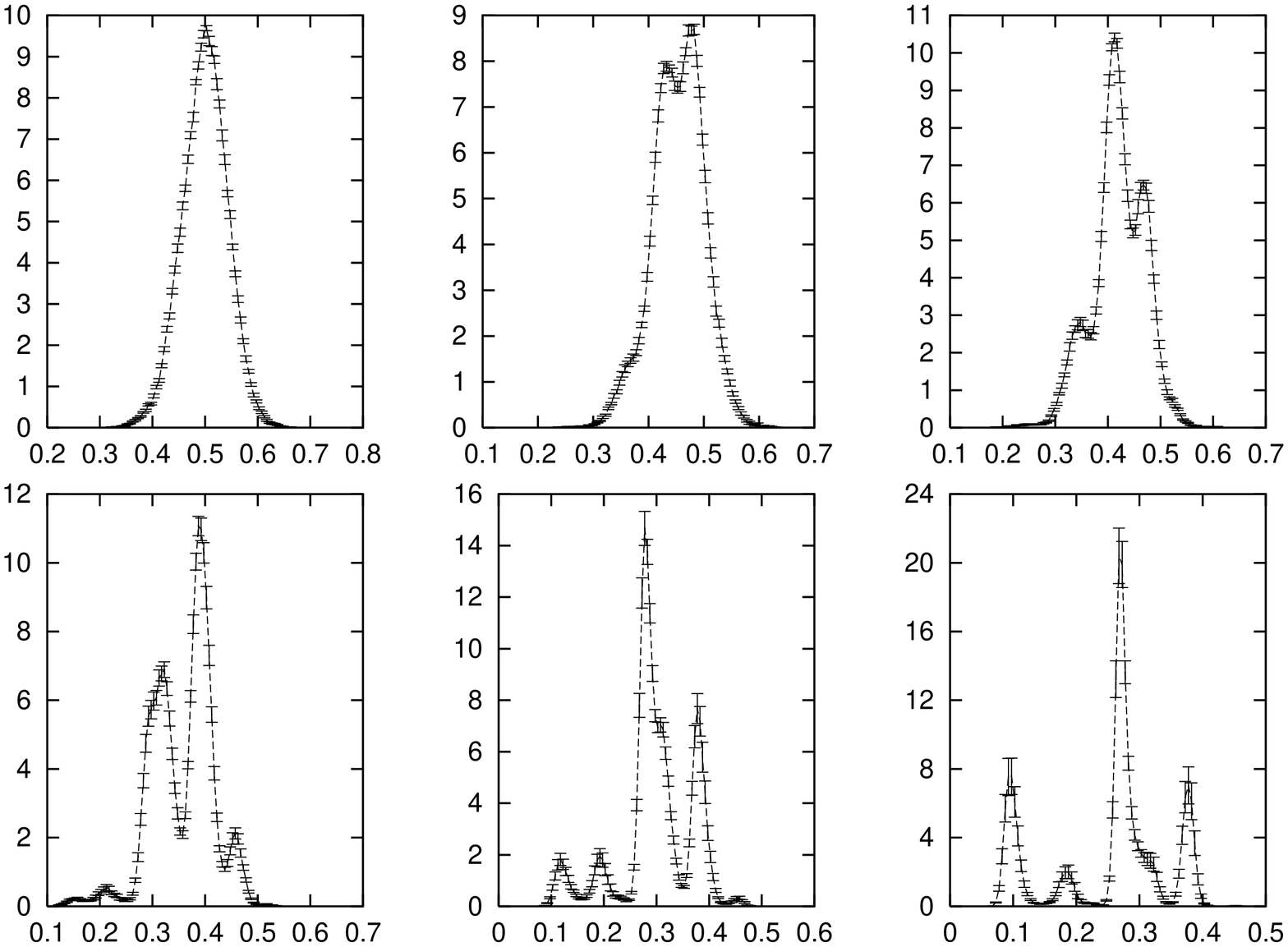,angle=270,width=10cm}} 
\caption{Data on $P(d)$ for $N$=36.  From left to right and top to bottom 
$\Gamma$=1.2, 1.3, 1.4, 1.6, 1.8 and 2.0 (i. e. the same value that in 
previous [Fig. 15]).}
\end{center}
\label{fig:15}
\end{figure}

Both the value of $\Gamma^{*}$ and the $P(d)$ shape in the glassy phase
are strongly dependent on $N$. Differences are evident also when the number
of particles is varied only of 2 (i.e. between $N$=28 and 30 or $N$=34 and 
36). This is easily understandable since a little difference in $N$ can 
change abruptly the kinds (and the number) of configurations that maximize 
and that are near to maximize relative distances between particles (minimizing 
the Hamiltonian). On the other hand, the observed non trivial dependence of 
the behaviour of the model on the number of particles seems to agree with 
$N$ playing in some way in structural glasses the same role that quenched 
random variables do in spin glasses. 

As a last remark we note that when reaching $\Gamma$ values well higher
than $\Gamma^{*}$ some peaks of $P(d)$ disappear and the height of the other 
ones (usually corresponding to smaller $d$ values) goes up. This seems quite
reasonable in a finite size system with continuous degrees of freedom since
in spite of the absence of a perfectly crystalline ground state only a few 
configurations of the particles are expected to really minimize the 
Hamiltonian.

The natural next step in this analysis should consist in trying to get the 
behaviour of quantities averaged over different $N$ values. We advance that 
from early new results \cite{cp} a slightly different definition of distance 
seems to be more suitable for mixtures and it could make easier this kind of 
study, permitting to gain a further insight into structural glass properties.

\section* {Acknowledgments} We acknowledge interesting discussions with
L. Angelani, A. Cavagna, S. Franz, I. Giardina and G. Ruocco.

\end{document}